# 80-degree field-of-view transmissive metasurface-based spatial light modulator.

## Authors

Anton V. Baranikov,[1*#] Shi-Qiang Li,[1#] Damien Eschimese,[1] Xuewu Xu,[1] Simon Thiele,[2] Simon Ristok,[3] Rasna Maruthiyodan Veetil,[1] Tobias W. W. Mass[1], Parikshit Moitra,[1] Harald Giessen,[3] Ramón Paniagua-Domínguez[1]*, Arseniy I. Kuznetsov[1]*

## Affiliations

[1]Institute of Materials Research and Engineering, A*STAR (Agency for Science, Technology and Research), 138634, Singapore

[2]Institut für Technische Optik and Research Center SCoPE, University of Stuttgart, Germany

[3]4th Physics Institute and Research Center SCoPE, University of Stuttgart, Germany

*Correspondence should be addressed to A.V.B. (Anton_Baranikov@imre.a-star.edu.sg), R. P.-D.(Ramon_Paniagua@imre.a-star.edu.sg) and A.I.K. (Arseniy_Kuznetsov@imre.a-star.edu.sg)

[#] These authors contributed equally to this work

## Abstract

Compact, lightweight and high-performance spatial light modulators (SLMs) are crucial for modern optical technologies. The drive for pixel miniaturization, necessary to improve their performance, has led to a promising alternative, active optical metasurfaces, which enable tunable subwavelength wavefront manipulation. Here, we demonstrate an all-solid-state programmable transmissive SLM device based on Huygens dielectric metasurfaces. The metasurface features electrical tunability, provided by mature liquid crystals (LCs) technology. In contrast to conventional LC SLMs, our device enables high resolution with a pixel size of ~1 μm. We demonstrate its performance by realizing programmable beam steering, which exhibits high side mode suppression ratio of ~6 dB. By complementing the device with a 3D printed doublet microlens, fabricated using two-photon polymerization, we enhance the field of view up to ~80º. The developed prototype paves the way to compact, efficient and multifunctional devices for next generation augmented reality displays, light detection and ranging (LiDAR) systems and optical computing.

**Introduction**

Recent advances in modern optical technologies, such as near-eye displays, free-space optical communications, digital holography, light detection and ranging (LiDAR) and optical sensing, have raised the demand for active devices being able to manipulate the wavefront of optical beams dynamically with high spatial resolution. One of the core elements for this purpose are spatial light modulators (SLMs) (1-4) capable to control the phase and the amplitude of the reflected or transmitted light on the pixel-by-pixel basis. Majority of commercially available SLMs employ liquid crystals (LCs) (5) or micro-electro-mechanical systems (MEMS) (6). The former remains dominant, while MEMS-based devices are less suitable for mass production due to high mechanical failure rates and challenging fabrication process (5,7). Typically, LC SLM functionality relies on the phase modulation provided by the reconfigurable LCs birefringence. Namely, the phase retardation of an incident polarized light is altered by LC molecules orientation, which in turn can be controlled by an applied electric field on the pixel-by-pixel basis. Though LC technology made a huge progress, reducing the LC SLM pixel size still remains challenging. The reason for this is the severe interpixel crosstalk at small pixel pitch, induced by the fringing fields, elastic forces in LC material and lateral ion migration (8-11). In turn, this dramatically limits the field of view of the LC SLM devices, defined as double the maximum deflection angle of the first diffraction order.

Optical metasurfaces have recently witnessed dramatic progress owing to their exceptional abilities in the light wavefront manipulation. Metasurfaces are planar arrays of optical nanoantennas providing subwavelength control over phase, amplitude and polarization of light (12-15). They have proven themselves as ultracompact, lightweight, highly efficient and multifunctional optical devices promising to replace conventional bulk optics. Starting from the static elements, such as waveplates (16), lenses (17-20) and polarimeters (21,22), metasurfaces have expanded into the realm of tunable devices (23-25), including realization of lenses with tunable focus (26-31), optical modulators (32,33) and beam steerers (30,34,35,36). The tunability is usually accomplished by altering the dielectric function of nanostructures via application of different external stimuli (electrical (30,37,38), thermal (39), mechanical (35), chemical (40), optical (41,42,43), etc.), or by changing the nanoantenna environment (27,29,44). Among the latter, interfacing nanoantennas with LCs shows particular promise to realize efficient devices operating at visible wavelengths, owing to their large permittivity modulation and transparency in this frequency range (29,44-47).

In this regard, recently, a proof-of-concept transmissive SLM based on electrically tunable Huygens dielectric metasurface has been reported (44). Here, elaborating on this concept, we present a fully programmable nanoantenna-based SLM with a large aperture comprising 96 individually addressable electrodes. For that, we integrate dielectric nanoantennas into a pixelated LC SLM and utilize the LC reconfigurable birefringence to tailor the metasurface phase response. Strong Mie-type resonances and Huygens condition enable achieving $2\pi$ phase modulation in discreet steps with high associated transmission, a must-have characteristic for a multifunctional SLM. With a pixel size of ~1 μm only and large pixel counts, this device surpasses previously reported device in terms of efficiency, which reaches ~30% experimentally in a dynamic beam steering configuration. Moreover, by leveraging on state-of-the-art 3D printing technology (48,49), we further expand the field-of-view of the metasurface SLM to nearly 80º, using an integrated doublet microlens fabricated via two-photon polymerization.

**Results**

*Design and nanofabrication of the metasurface SLM*

We design the dielectric metasurface based on TiO$_2$ nanopillars providing low absorption losses in the visible spectral range (50). Owing to their high refractive index (~2.5), such structures support strong localized Mie-type magnetic and electric dipolar resonances, whose spectral positions can be tailored by geometry (51-53). The nanopillars are encapsulated in a nematic LC environment (QYPDLC-001C, extraordinary index $n_e$=1.81, ordinary index $n_o$=1.52, layer thickness ~1.3 µm) sandwiched between a set of 96 linear, pixelated bottom electrodes and a uniform top electrode, all made of indium tin oxide (ITO) on quartz. Each pixel contains 3 nanopillars in the transverse direction (see Fig. 1A). The distance between the pillars within one pixel is chosen to be sub diffractive (360 nm), while the electrodes pitch is 1.14 µm. In the absence of an applied voltage, the LC director is oriented in-plane. The device is illuminated by a normally incident beam with the electric field polarized parallel to the LC director, so that it experiences the extraordinary refractive index. The voltage application induces a reorientation of the LC molecules (Fig. 1B), which alters the nanoantennas environment refractive index. This, in turn, enables a spectral shift of the resonances. By bringing the electric (ED) and magnetic dipoles (MD) into spectral overlap at a certain LC rotation, the Huygens' condition is satisfied, resulting in 100% transmission and full $2\pi$ phase coverage (54-56).

To optimize the nanoantennas dimensions, we perform numerical simulations (see Methods) and obtain the transmission and phase spectra of the metasurface for different LC molecules orientation (Fig 1C and 1D). For the nanopillars with a height of 195 nm and a diameter of 270 nm, we observe spectrally separated MD (~659 nm) and ED (~675 nm) resonances for in-plane LC orientation. Upon the rotation of LC molecules, the resonances start to approach each other and, when they overlap, the Huygens' condition is met at around 672 nm for ~50-degree LC director orientation. As can be seen, when the condition is met, transmission values >60% are achieved in the whole resonance overlapping region, with associated ~$2\pi$ phase modulation. It is worth noting here that only distinct phase values are accompanied with high transmission. Consequently, the metasurface supports discrete phase levels, namely two or three, depending on the device configuration, which span the $2\pi$ range (for the detailed discussion see Supplementary Materials, Section 1).

We fabricate the designed metasurface SLM, containing 96 individually addressable bottom electrodes, using a combination of double-layer electron beam lithography (for the nanoantenna and bottom electrode patterning) and photolithography (to fan out the electrodes and create the bond pads for wire bonding). Fig. 1E shows the photographic image of the sample mounted on the driving printed circuit board (PCB) together with the microscope image of the device. The electrodes are arranged in the butterfly shape, routing them to the rectangular metasurface active area (120x100 µm$^2$) containing the nanopillars. Scanning electron microscope (SEM) images of the metasurface before LC infiltration are presented in Fig.1F. For specific details of the device fabrication and the electrodes addressing see Methods and Supplementary Movie 1.

*Optical performance of the metasurface SLM*

We characterize the performance of the fabricated SLM device by realizing programmable beam steering. As a first validation, we implement simple binary gratings in which the grating pitch is reconfigured. To do so, two alternating groups of electrodes are either kept grounded or biased to induce a rotation of the LC (with typical biases in the range of 2-3V, as described below). By changing the number of electrodes in these groups, a different grating pitch (p) is obtained and the diffraction angle is changed (see insets in Fig. 2A). The target for the grounded electrodes is keeping the LC director in-plane (0 degree) while the target for the biased ones is achieving ~50 degree out-of-plane LC orientation. Note that the latter state exploits the Huygens' condition, which ensures the necessary phase modulation (~$\pi$) and a low associated transmission modulation.

For the optical characterization, we implement spectrally resolved back focal plane (BFP) measurements (see Methods). A collimated broadband light illuminates the device and the diffraction pattern provided by the SLM in transmission is collected by an objective, whose BFP is subsequently coupled to a spectrograph equipped with a CCD camera. In this way, spectral and angular information of the far-field intensity profile is obtained. Fig. 2A depicts the measured far-field angular intensity distribution for the cases of no applied voltage (top panel) and 6 different grating periods. A colored pattern in the inset of each panel depicts the corresponding voltage pattern.

In the case of no applied voltage, we observe, as expected, pure $0^{th}$ order transmission, owing to the sub-diffractive nanoantenna period. Applying the voltages, we observe the emergence of $+1^{st}$ and $-1^{st}$ diffraction orders, with deflection angles perfectly matching the theoretical ones given by $\alpha = \pm\sin^{-1}(\lambda/p)$ (depicted in Fig.2A as grey dashed vertical lines). Note that, for each case, an optimum operational wavelength is chosen as the one maximizing the ratio $(I_{+1} + I_{-1})/I_0$, where $I_{+1}$, $I_{-1}$, $I_0$ are the intensities of $+1^{st}$, $-1^{st}$ and $0^{th}$ orders, respectively. As an example, the diffraction efficiency spectra into these three orders are shown in Fig. 2B, when the device is configured as a grating with 2 electrodes for each phase level. The red vertical line indicates the optimum wavelength (around 651 nm), for which ~30% of the incident power is channeled into the $\pm1^{st}$ orders, while the $0^{th}$ order remains at ~5% only. For benchmarking, the simulated results (computed assuming perfect in-plane orientation of the LC on the unbiased electrodes and 50-degree LC rotation on the biased ones) are shown in Supplementary Materials, Fig. S1. In both simulations and experiment, one may notice an asymmetry between the $+1^{st}$ and $-1^{st}$ orders. In the former case, unidirectional rotation of the LC molecules across the whole device induces slight variations in dielectric permittivity tensor for the $1^{st}$ and $-1^{st}$ order propagation directions (57). In turn, this leads to the different diffraction efficiencies. In the experiment, two more reasons can contribute to the asymmetry: fabrication imperfections (see Supplementary Materials, Fig. S3) and an existing pretilt angle (relative to the metasurface plane) of the LC molecules. The latter one is conventionally done on purpose in LC devices to force unidirectional rotation upon the voltage application. Combined with the fringing fields, it leads to the asymmetric phase profile and was shown to cause inequal $+1^{st}$ and $-1^{st}$ diffraction order intensities for LC binary gratings (58, 59). In our device, an alignment polyamide layer is placed on top of the LC layer, causing a pretilt angle. Moreover, the bottom LC-nanopillars interface may also lead to an additional out-of-plane alignment.

The most striking result to emerge from the far field distributions is the strong $0^{th}$ order suppression starting from the case of 2 electrodes per each phase level. The Side Mode

Suppression Ratio (SMSR), determined as the intensity ratio between the main mode (+1$^{st}$ or -1$^{st}$ order) and the largest side mode (0$^{th}$ order), is found to be around 6 dB, a remarkable value for a multi-electrode tunable metasurface. One can notice that the smallest grating pitch (the largest deflection angle) does not provide as good 0$^{th}$ order suppression as the other cases. This is attributed to the interpixel LC crosstalk effects such as fringing fields, elastic forces and lateral transport of ionic impurities in the LC that, although minimized, are still present in this thin LC cells. The first two prevent abrupt phase variations, acting as a low pass filtering of an ideal phase profile (8,9), while the lateral ion transport gives rise to electric field screening and perturbations in LC alignment (10,11). All these effects influence the designed phase profile and become more prominent for a smaller pixel dimension. To corroborate this, we plot the diffraction efficiency (defined as the total transmission intensity of the +1$^{st}$ and -1$^{st}$ orders normalized to the incident light) in Fig. 2C. As seen there, upon increase of the deflection angle, the efficiency gradually decreases, starting from near 40% down to ~10%.

To further test the functionality of our SLM, we operate the device as a programmable, gradient blazed grating, introducing three voltage levels: a grounded one and two elevated ones (hereafter denoted as *medium* and *high* voltage). In this situation, the incident power is expected to channel into a single diffraction order, thus operating in a beam bending, rather than a beam splitting, regime. Fig. 3A presents the far-field patterns for several grating periods and the corresponding voltage configurations (shown as colored patterns in the inset of each panel). For each case, the wavelength and the voltage levels are optimized according to the optimization procedure explained below, leading to strongly asymmetric angular intensity distributions with light deflected prominently into the +1$^{st}$ order and confirming the correct phase gradient distribution introduced by the metasurface SLM. Similarly, to the beam splitting regime, the smallest supercell pitch (corresponding to the largest bending angle ~11°) provides the worst SMSR value and the worst bending efficiency of ~10% (see Fig.3B). For smaller bending angles, however, SMSR is calculated to be in the range of 4-5.5 dB, with bending efficiencies (normalized to the incident light) gradually increasing from ~15% to ~30%. It is worth noting that, due to the thin character of the LC cells, the required voltage levels are quite small and these need to be carefully optimized. Fig. 3C presents the voltage optimization study for the 2 electrodes per phase level case, where we plot the normalized directivity into the desired order (defined as the bending order intensity normalized to the integrated background across the whole angular range, including all other diffraction orders, i.e. $I_{+1}/\sum_i I_i$ for all *i*) as we vary the *medium* (0.5-2.5 V) and *high* (2.0-4.5 V) voltage levels. One can identify three optimum voltage combinations, namely, (1.0;2.5), (1.5;2.5) and (1.0;3.0), all leading to good directivity. Additionally, we study the inverse beam bending regime, where the blazed grating phase profile is mirrored and deflects the light into the opposite -1st order. We note, however, that the obtained SMSR value is noticeably lower in this case, most probably due to the same reasons responsible for the asymmetry in the diffraction orders in the beam splitting regime discussed above.

*The metasurface SLM field of view enhancement*

While the field of view of the presented device is many-fold larger than any commercial SLM, it is still insufficient for certain applications (e.g. LIDAR). In this section, we combine our metasurface SLM with a 3D printed microlens doublet, as to further expand the field of view to ~80°. The doublet is designed using commercial ray-tracing software (Zemax OpticStudio) to provide 5× angular expansion of the SLM diffraction pattern. Note that, being cylindrical, the doublet acts along one spatial axis only. Fig.4A presents the ray tracing diagram, where each specific color corresponds to a particular angle and the input aperture of the doublet allows for ~ (-10º,10º) input angles. We fabricate the doublet using the two-photon polymerization assisted 3D printing technique, which enables arbitrary free-form shapes and precise alignment between the optical components (48,49). The detailed fabrication process of the doublet microlens is discussed in Methods. To test the performance of the combined SLM-microlens doublet system, we perform BFP measurements using a coherent laser source (for the optical setup design see Supplementary Materials, Section 3). For the test, we operate the device in the beam splitting regime, since it provides the largest deflection angle and better SMSR ratio. Fig. 4B summarizes the measurements results. First, we repeat the measurements for the SLM only, i.e. without the integrated microlens (Fig.4B, the left panel), except of the 1 electrode per phase level case, for which the steering angle falls outside the acceptance angle of the doublet. Note that, compared to the results shown in Fig. 2B, the $0^{th}$ order is slightly larger here, due to the narrow-band spectral response of our metasurface and finite bandwidth of the laser (full width at half maximum ~2 nm). Next, we integrate the microlens doublet and perform the measurements for the same electrode configurations. To do so, we use an immersion oil on the interface between the SLM superstrate and the doublet microlens substrate, to match the refractive indices and avoid parasitic Fabry-Perot modes. The right panel of Fig. 4B presents the obtained results, clearly showing the diffraction angle expansion, which results in a total field of view of ~80º. One might notice the wider far field spots in comparison with the SLM alone, which follows from the additional beam divergence provided by the doublet (for the detailed discussion see Supplementary Material, Section 3). We also observe a slight nonuniformity in the measured magnification dependence on the input angles (Fig. 4C), spanning from 3.5X for the lowest one (~3º) to 4.5X for the largest one (~8.5º). This can be explained by the non-ideal microlens shape and the additional immersion oil layer, which is not taken into account in the design. For the photographic images of the combined device and a magnified view of the microlens doublet see Fig. 4D. The interested reader can refer to Supplementary Movies 2 and 3 for a video recording of the device steering a beam in real-time without and with the integrated doublet, respectively. In these demonstrations, the laser is focused to match the size of the active area by a lens with 200 mm focal length. The far-field distribution is visualized on a white paper card.

**Discussion**

In conclusion, we have designed, fabricated and characterized a tunable metasurface transmissive SLM device with 96 individually addressable electrodes. The nanoantennas comprising the metasurface provide the necessary phase accumulation and allow to reduce the thickness of LC layer. This, in turn, reduces interpixel crosstalk and helps to shrink the pixel size down to 1.14 μm. By programming the individual electrodes, we have realized dynamic beam steering in two regimes, namely, beam splitting and beam bending (with the device acting as programmable binary and blazed diffraction gratings, respectively). The SLM shows high-performance, reaching 6 dB SMSR for a field of view of 17º. Maximum efficiencies ~40% and ~30% are achieved, respectively, for the beam splitting and beam bending configurations, with a gradual drop to ~20% and 15%, respectively, upon the increase of the diffraction angle. To further expand the device FOV, we complement it with a 3D printed microlens doublet designed to create a 5X angular magnification. This allowed us to increase the field of view up to ~80º.

Our prototype performance might be further improved by state-of-the-art artificial intelligence (AI) to mitigate any residual interpixel crosstalk effects (60). Furthermore, AI was shown to overcome covarying phase and amplitude modulation inherent for metasurfaces, which is also present, to a certain extent, in this device (61). We also expect that future research will be concentrated on multifunctional metasurfaces with individually addressable 2D pixel arrays. High SMSR, the absence of mechanical parts, thus enhancing the device reliability, the small pixel size, low driving voltage and flexible tunability make our device promising for a variety of applications, ranging from LiDAR to tunable imaging and real-time holography, to mention some.

**Materials and Methods**.

*Fabrication*

The device is fabricated on a commercial 20x20mm glass slide coated with a 23nm ITO film (Latech Scientific Supply Pte. Ltd, Singapore), which is used to create the bottom electrode pattern. On top of it, a 200nm thick layer of amorphous $TiO_2$ is deposited via an ion-assisted deposition system (Oxford Optofab 3000) followed by a 30nm thick layer of chromium (Cr) deposited by evaporation (Evovac, Angstrom Engineering). To pattern the nanoantennas into the $TiO_2$ layer, the Cr is used as a negative hard mask. This mask is created using a spin-coated and baked layer of hydrogen silsesquioxane (HSQ, Dow Corning, XR-1541-002) exposed by electron-beam lithography (EBL, Elionix, 100 kV) and developed in Tetramethylammonium hydroxide solution (TMAH, 25%). The pattern is then transferred from the resist into the Cr using reactive-ion-etching (RIE) process (Plasmalab System 100, Oxford Instruments) with $Cl_2$ and $O_2$ gases, and from the Cr mask into the $TiO_2$ layer with $CHF_3$ gas via the same RIE system. The hard mask is then removed by Cr etchant solution (Sigma-Aldrich).

A second EBL process is performed to form the electrodes underneath the metasurface. A spin-coated and baked positive resist (ZEP520A, ZEONREX Electronic Chemicals) is used as a mask for the electrode patterning. The electrode design is transferred into the ITO layer via a similar RIE process using $CH_4$ and Ar gases. The whole ITO pattern on the substrate could not be made using only EBL due to a long writing time. Therefore, ITO connectors and bond pads, required for the wire-bonding step, are generated using photolithography and connected to the electrodes previously fabricated by EBL. For that, a spin-coated and baked positive photoresist (S1811, Microposit, ROHM AND HAAs) is exposed using a mask-aligner system (EVG6200, EV group). The connectors and bond pads patterns are then transferred to the remaining ITO layer through the same RIE recipe as before.

The ITO bond pads are covered with gold and connected with the electrical wires of the Printed Circuit Board. For that, 50nm of Cr, acting as an adhesion layer, is first deposited via evaporation following by 200nm of gold deposition using the same tool (Evovac, Angstrom Engineering). A second photolithography process through the mask-aligner system exposed another spin-coated and baked photoresist on the substrate to mask the bond pad geometry. Successive wet-etching processes are performed to transfer the pattern to the gold and the chromium layers (Sigma-Aldrich).

The device is finalized by assembling a second ITO-coated glass, serving as the common, top electrode, and the aforementioned metasurface with the bottom ITO electrode pattern to form a cell for liquid crystal filling. On top of the common electrode, polyamide is coated and rubbed, to define the liquid crystal molecular orientation in the absence of electric bias. The thickness of the assembled cell is defined by UV adhesive and silica spacers. The nematic liquid crystal QYPDLC-001C is encapsulated in the cell through capillary filling.

*The 3D printed lens doublet* (acting as cylinder lens) was designed using ZEMAX (Version 13). The surfaces are defined according to the Even Asphere model $z(r) = r^2/R \left(1 + \sqrt{1 - r^2/R^2}\right) + a_2 r^2 + a_4 r^4 + a_6 r^6$ with the parameters given in the table below.

| Interface | Distance to next surface (μm) | Radius of curvature R (mm) | $a_2$ (mm$^{-1}$) | $a_4$ (mm$^{-3}$) | $a_6$ [mm$^{-5}$] |
|---|---|---|---|---|---|
| Substrate | 112.9 | ∞ | - | - | - |
| Lens surface 1 | 351.0 | -0.989 | -1.531 | 4.715 | -17.474 |
| Lens surface 2 | 69.6 | -0.655 | -4.615 | 127.265 | -4538.658 |
| Lens surface 3 | ∞ | -0.568 | 2.122 | -36.356 | 178.906 |

The entire optical element has a size of 550 x 200 x 620 μm³. The 3D printing was carried out using a Nanoscribe GT system. As resist we used Nanoscribe IP-S resist. The writing was performed using a 25X objective with 200 nm slicing distance, 500 nm hatching distance, a scan speed of 50 mm/s and 70% laser power.
  Characterization was carried out with a Nanofocus confocal optical profiler and a Keyence VX 3D microscope. The laser wavelength is 780 nm and the pulse duration is 150 fs.

*Numerical simulations*

We used the finite difference time domain solver of Maxwell equations from Lumerical FDTD Solutions (62) to simulate the SLM performance. The results presented in Fig. 1 were obtained by simulating a single unit cell, with Periodic Boundary Conditions (PBC) applied in the transverse direction (thus mimicking an infinite system) and Perfectly Matched Layers (PML) in the top and bottom directions. An incident plane wave is injected from the top of the simulation domain and the transmitted and reflected fields recorded using appropriate 2D monitors. For the obtained beam steering results presented in Fig. 2, the TiO$_2$ nanoantennas were regularly placed on the bottom ITO electrodes with 3 nanoantennas per pixel. A supercell containing 2 pixels was considered, and PBC and PML boundary conditions, again, applied in the transverse and top and bottom directions, respectively. The diffraction efficiencies were computed using the built-in analysis group for *grating order transmission* calculation.

*Electro-optical characterization*

For the electrical control, a 96-channel 12-bit digital-to-analog converter (DAC) board (DAC60096EVM Texas Instruments) was used. To tailor the DAC output voltages independently, we used an ESP32 microcontroller (MCU from Espressif Inc), which was programmed in Python. We implemented AC voltage driving by external toggling of the DAC board with a waveform generator. This was done to avoid vertical ion migration effects in the LC layer, which may lead to the internal electric field screening (short-term) and the ion accumulation in the alignment layers (long-term) (10).

The optical performance of the metasurface SLM (Fig.2 and Fig.3) was characterized using spectrally resolved back focal plane (BFP) imaging (63). The sample was illuminated by a collimated broadband light (angular spread ~0.6º). The diffraction pattern, provided by the SLM in transmission, was collected by an objective (Nikon 20X, NA=0.45). Next, the BFP image was coupled to a spectrograph equipped with a CCD camera (Andor Kymera 328i). The spectrograph slit is aligned perpendicular to the metasurface electrodes, along the deflection direction. The angular resolution is ~0.2º, while the spectral one is ~0.6 nm. The image plane was spatially filtered to cut out the light transmitted outside of the active metasurface area.

For the combined SLM-doublet device characterization (Fig.4), we used a similar BFP imaging technique but using a coherent laser source instead (supercontinuum fiber laser SuperK EXTREME equipped with a tunable single line filter SuperK VARIA). For the setup details, see Supplementary Materials, Section 3 and Supplementary Materials, Fig.S4.

**Funding:** This work was supported by National Research Foundation of Singapore under Grant No. NRF-NRFI2017-01, IET A F Harvey Engineering Research Prize 2016, and AME Programmatic Grant No. A18A7b0058 (Singapore). We also acknowledge funding from BMBF (Printoptics and Exist), ERC (PoC 3D PrintedOptics, grant number 862549), BW-Stiftung (Opterial), DFG GRK2642, and MWK (ICM).

**Author contributions:** A.V.B. performed the electro-optical characterization and wrote the first draft. S.-Q. L. and D. E. performed the device nanofabrication, with help from P. M. R. M. V. performed the device packaging. X. X. carried out the numerical simulations and initial optical characterization. T. W. W. M. developed the electrical addressing protocol. S. T. designed the microlens doublet and S. R. fabricated it. H. G., R. P-D. and A. I. K. conceived the idea and coordinated the research. All authors analyzed the data and read and corrected the manuscript.

**Competing interests:** Authors declare that they have no competing interests.

**Data and materials availability:** All data are available in the main text or the supplementary materials.

**Figures**

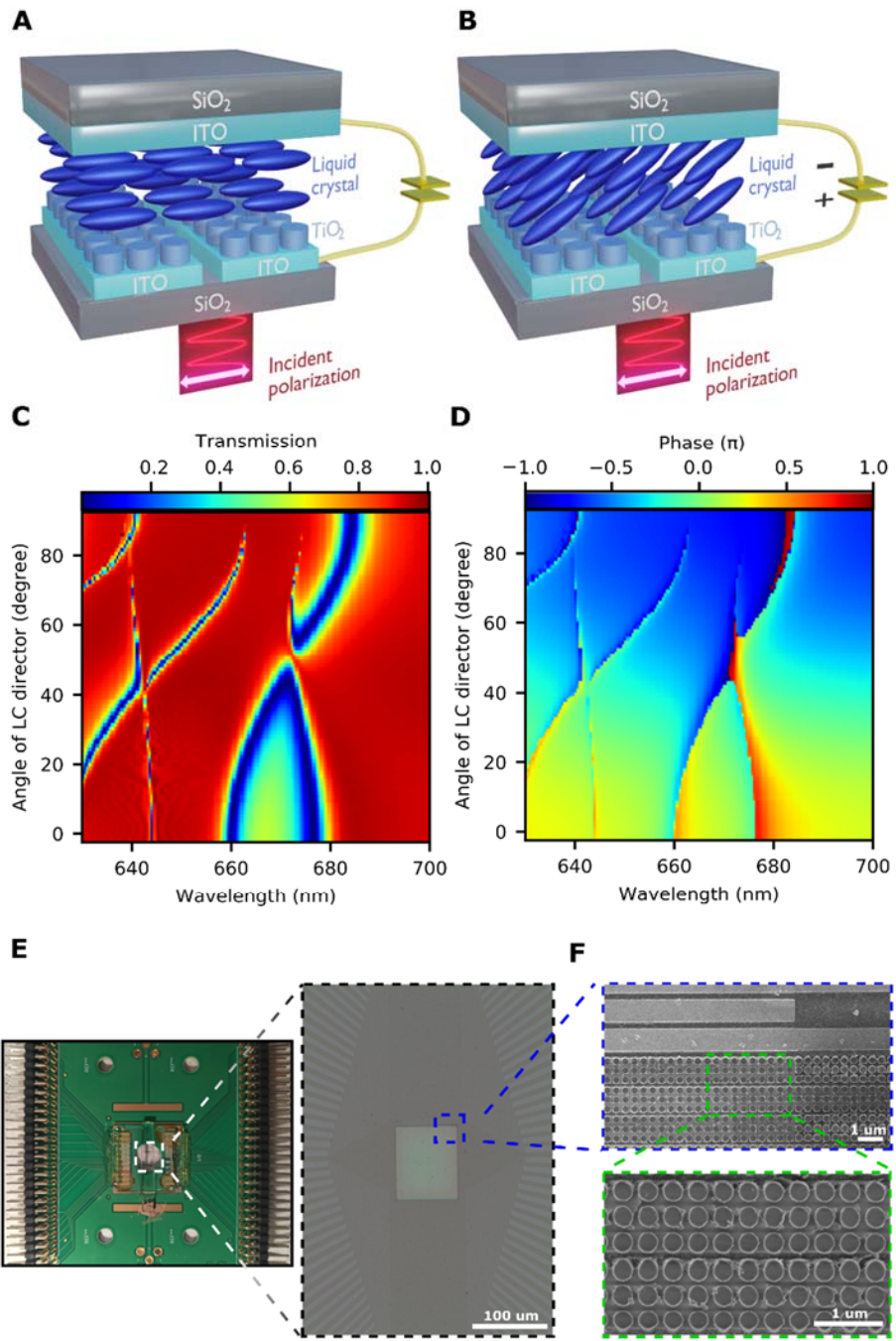

**Fig. 1. The transmissive metasurface SLM.** (A) An artistic-view, schematic of the SLM architecture. The nanopillars are embedded in the LC layer sandwiched between the (linear) bottom pixelated and uniform top ITO electrodes. Two electrodes are shown here. Each electrode (pixel) contains 3 nanopillars in the transverse direction. In the absence of applied voltage, the normally incident, linearly polarized light experiences the extraordinary refractive index of the LC. (B) Externally applied voltage induces LC molecules reorientation, altering the refractive index experienced by the impinging light and surrounding the nanopillars. (C-D) Simulated transmission (C) and phase (D) spectra of the metasurface for different LC molecules orientation. The angle of the LC director is taken relative to the metasurface plane. The Huygens condition is met at around 672 nm at 50-degree LC

director orientation (E) A photographic image of the PCB (left panel) together with the magnified optical microscope image of the metasurface active area (right panel). The electrodes are arranged in the butterfly shape. (F) SEM images of the metasurface before LCs infiltration, showing the nanopillars and pixelated electrodes.

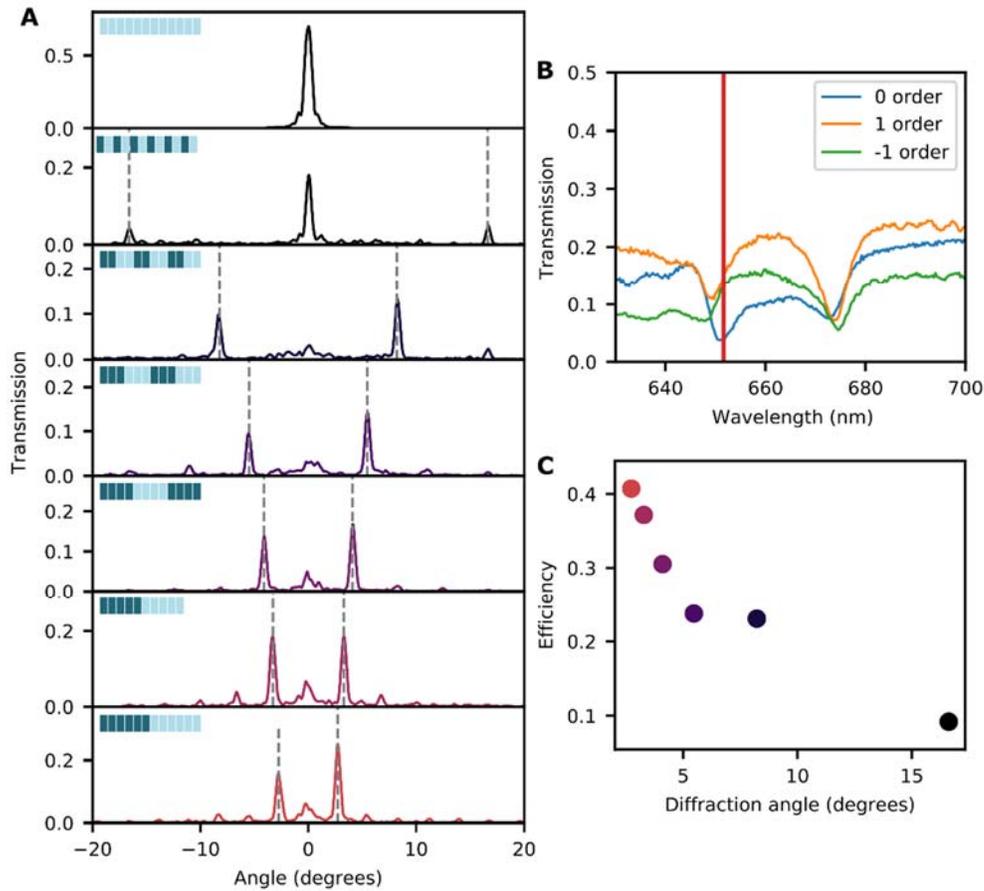

**Fig. 2. The metasurface SLM beam steering in the beam splitting regime.** (A) Experimental far-field angular intensity distribution for the cases of no applied voltage (top panel) and 6 different binary grating periods. The inset of each panel schematically depicts the corresponding voltage pattern. The vertical, gray dashed line indicates the theoretical deflection angles. (B) Measured transmission spectra of the +1$^{st}$, -1$^{st}$ and 0$^{th}$ diffraction orders (orange, green and blue curves) in the case of 2 electrodes per each phase level. The red vertical line indicates the optimum wavelength. (C) The diffraction efficiency versus diffraction angle. The efficiency is defined as the total intensity of the +1$^{st}$ and -1$^{st}$ orders normalized to the incident light.

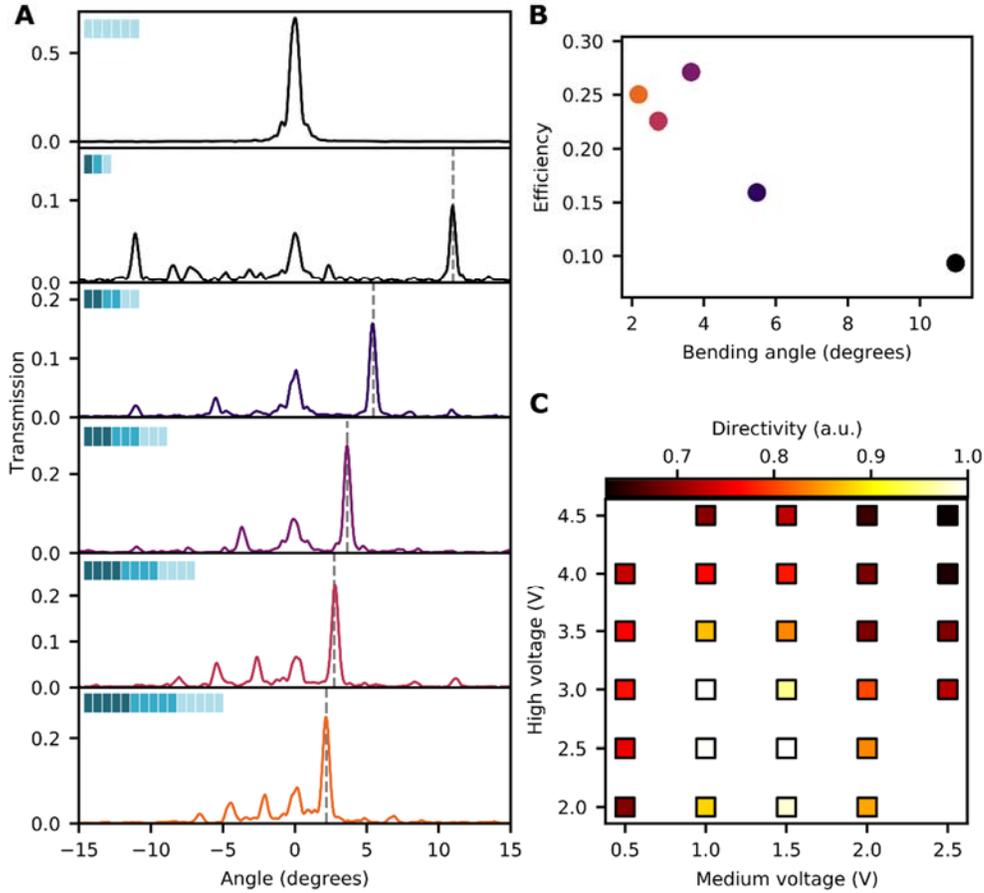

**Fig. 3. The metasurface SLM in the beam steering regime.** (A) Experimental far-field angular intensity distribution for the cases of no applied voltage (top panel) and 5 different blazed grating periods. The inset on each panel schematically depicts the corresponding voltage pattern. The vertical, gray dashed lines indicate the theoretical deflection angle. (B) Bending efficiency (defined as the intensity of the +1$^{st}$ order normalized to the incident light) versus bending angle. (C) Voltage optimization study for the case of 2 electrodes per phase level. The intensity plot shows the bending directivity (defined as the bending order intensity divided by the integrated background across the whole angular range including all other diffraction orders: $I_{+1}/\sum_i I_i$ for all $i$) for various combinations of medium and high voltages. The directivity is given in arbitrary unites, normalized to its own maximum.

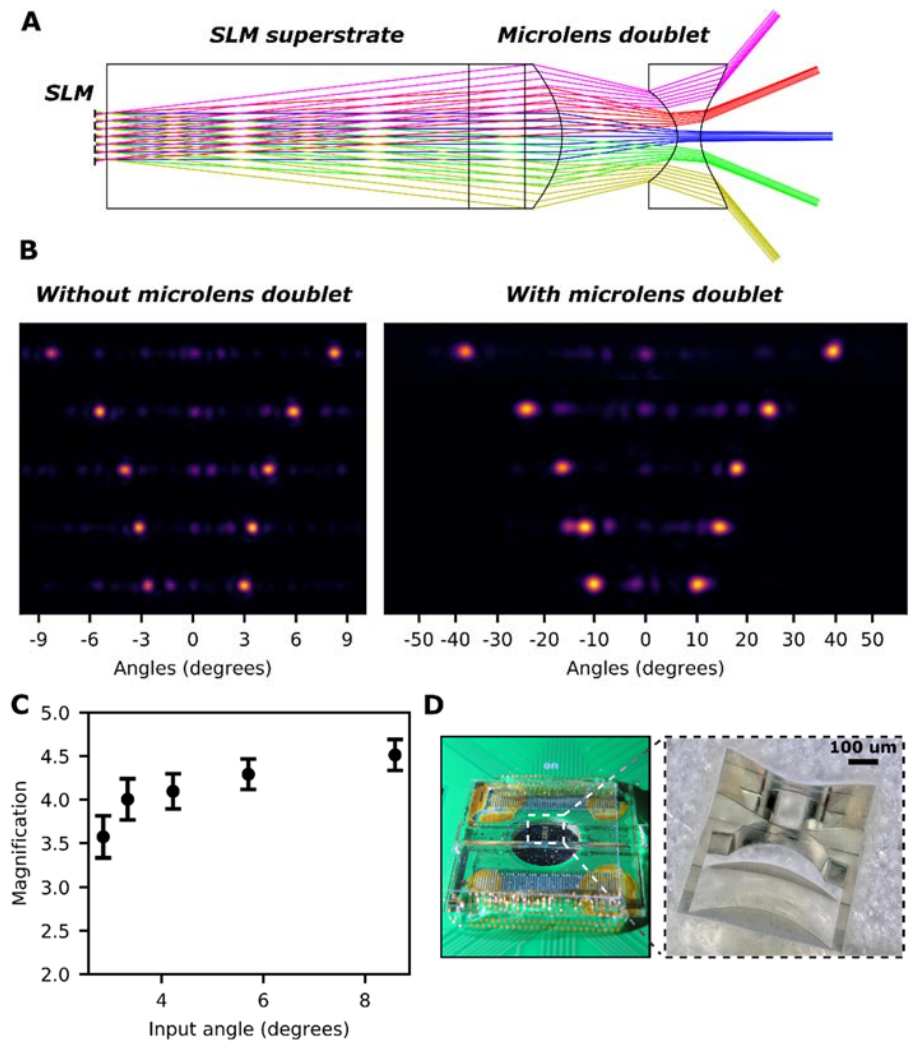

**Fig. 4. The metasurface SLM with enhanced field of view.** (A) Ray tracing diagram of the SLM with the integrated microlens angular magnifier. Each color corresponds to a particular steering angle. (B) Far-field intensity distributions in the beam splitting regime for 5 steering angles. The left panel depicts the results obtained for the SLM without the integrated microlens doublet, while the right panel shows the results for the combined SLM-microlens doublet configuration. (C) Measured angular magnification as a function of input angle. (D) Photographic images of the combined SLM-microlens doublet configuration (left) and a magnified view of the doublet lens (right).

# Supplementary Materials for

## 80-degree field-of-view transmissive metasurface-based spatial light modulator.

Anton V. Baranikov*, Shi-Qiang Li, Damien Eschimese, Xuewu Xu, Simon Thiele, Simon Ristok, Rasna Maruthiyodan Veetil, Tobias W. W. Mass, Parikshit Moitra, Harald Giessen, Ramón Paniagua-Domínguez*, Arseniy I. Kuznetsov*

*Corresponding authors. Email: Anton_Baranikov@imre.a-star.edu.sg, Ramon_Paniagua@imre.a-star.edu.sg, Arseniy_Kuznetsov@imre.a-star.edu.sg

Section 1. Three-level phase-only metasurface SLM based on Huygens' condition

In this Section, we show how to use the Huygens' condition to achieve discrete level phase control with small coupled amplitude modulation for the beam steering realization. Figure S1A shows the simulated metasurface phase (light blue curve) and transmission (light red curve) versus the LC rotation at a particular wavelength (670.5 nm) in the vicinity of the Huygens' condition. As discussed in the main text, we target 0- and 50-degree LC orientation to implement the beam splitting. Indeed, these states (denoted as black dashed lines in Fig. S1A) correspond to the necessary phase values (0 and ~$\pi$, see Fig. S1A) and moderate transmission modulation (49.3 % and 71 %, respectively). We utilize the two states to construct an infinite binary grating with a grating pitch of 2.28 μm (2 electrodes) and simulate the diffraction of a normally incident linearly polarized light (see the main diffraction orders spectra in Figure S1B). One can observe a clear beam splitting with $0^{th}$ order suppression and ~60% total diffraction efficiency around 674 nm.

Beam bending regime can be similarly implemented by adding a third phase level, as to construct a blazed grating with ~$2\pi/3$ phase steps. For this, the optimum wavelength is slightly redshifted towards 672 nm (see Fig. S1C for the simulated metasurface phase, indicated as a light blue curve, and transmission, denoted by a light red curve, versus the LC rotation). The black dashed lines indicate the three utilized states, having 0, ~$0.68\pi$, ~$-0.78\pi$ phases and 37.2 %, 82.8 %, 96,8 % transmission values. As expected, such blazed grating results in funneling of the optical power into a single diffraction order (see the simulated beam bending in Fig.S1D) with ~40% bending efficiency around 672 nm.

It should be noted that there is an intrinsic limitation in this device design concerning the number of discrete phase levels achievable. One can notice that the phase jumps occur together with certain transmission modulation, which is inherent to resonance shifts (see, for example, Fig.S1C in the 40-60 degrees range of the LC rotation). Essentially, this leads to a larger transmission modulation upon increasing the number of phase steps and deteriorates the device performance.

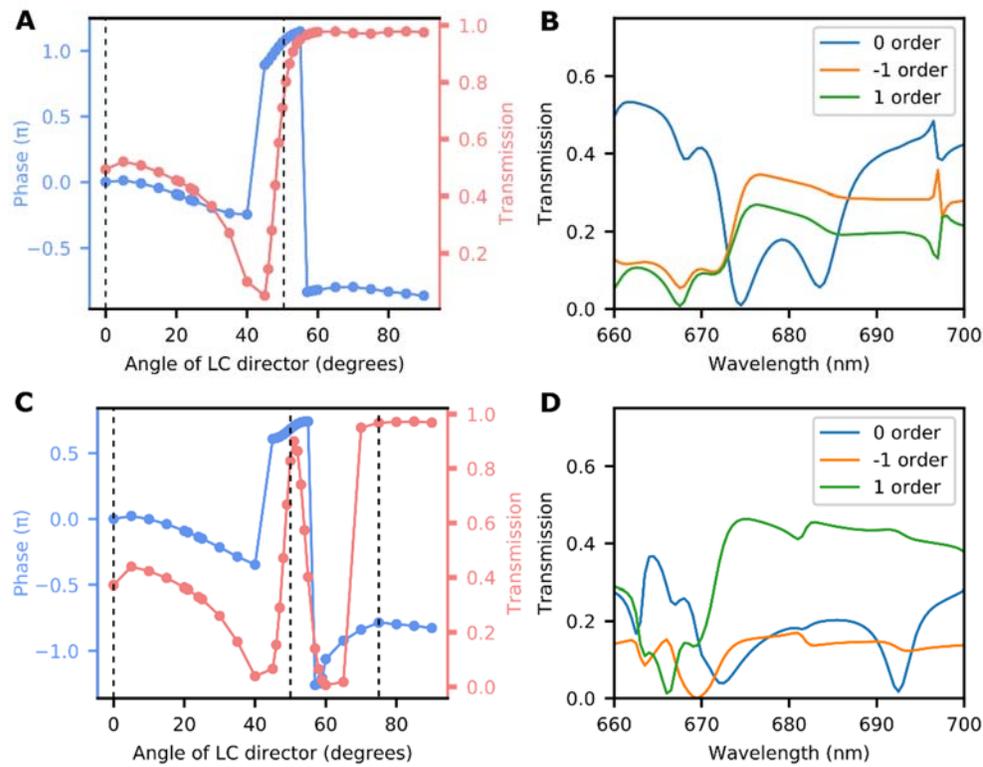

**Fig. S1.**
**Using Huygens' condition for the discrete phase control. (A)** Simulated metasurface phase (light blue curve) and transmission (light red curve) versus the LC rotation at a particular wavelength (670.5 nm) in the vicinity of the Huygens' condition. The black dashed lines indicate the two states used for the beam splitting. **(B)** Numerical simulation of the beam splitting. **(C)** Simulated metasurface phase (light blue curve) and transmission (light red curve) versus the LC rotation at a particular wavelength (672 nm) in the vicinity of the Huygens' condition. The black dashed lines indicate three states used for the beam bending. **(D)** Numerical simulation of the beam bending.

Section 2. Inverse beam bending regime

In this Section, we provide the measurements of the inverse beam bending regime, i.e. applying the blazed grating phase profile mirrored with respect to the one presented in the main text. As discussed therein, the wavelength and the voltages are optimized for each grating supercell. Fig.S2 compares the angular intensity distribution for the inverse beam bending (top panels, blue curves) and the beam bending regime shown in the main text (bottom panels, green curves). Fig.S2A, S2B and S2C depict the cases of 1,2 and 3 electrodes per phase level, respectively. As observed, the inverse beam bending regime doesn't provide as good SMSR value. We attribute this to the same reasons that induce the asymmetry between the diffraction orders in the beam splitting regime: the unidirectional rotation of the LC molecules across the whole device, the LC pretilt angles (see the discussion in the main text) and fabrication imperfections. To illustrate the latter, we show a magnified SEM image of the metasurface (Fig. S3). There, one can see that the bottom ITO electrodes are slightly truncated (see the dashed black line indicating such an example for one of the electrodes). This leads to a relative shift between the electrode positions and those of the 3-nanopillar supercells.

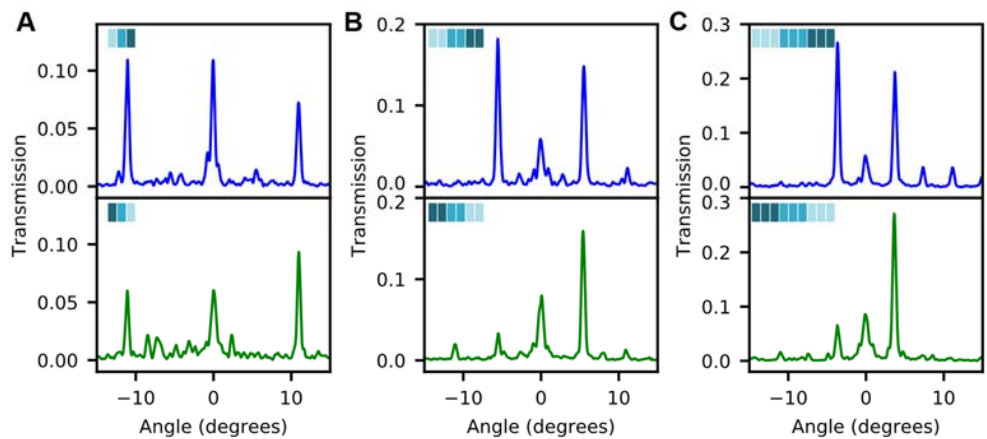

**Fig. S2.**
**Inverse beam bending regime.** **(A)** The angular intensity distribution for the case of a blazed grating with 1 electrode per phase level in the inverse beam bending regime (top panel, blue curve) and the beam bending regime presented in the main text (bottom panel, green curve). **(B)** The angular intensity distribution for the case of a blazed grating with 2 electrodes per phase level in the inverse beam bending regime (top panel, blue curve) and the beam bending regime presented in the main text (bottom panel, green curve). **(C)** The angular intensity distribution for the case of a blazed grating with 3 electrodes per phase level in the inverse beam bending regime (top panel, blue curve) and the beam bending regime presented in the main text (bottom panel, green curve)

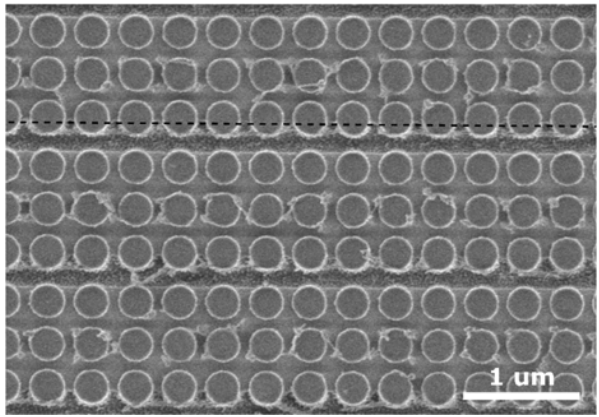

**Fig. S3.** Magnified SEM image, showing a slight displacement between the ITO electrodes and the 3 nanopillars supercells.

Section 3. Optical characterization of the metasurface SLM combined with the doublet microlens

In contrast to the broadband back focal plane (BFP) measurements presented in Fig. 2 and Fig. 3 of the main text, we utilize a coherent laser source (supercontinuum fiber laser SuperK EXTREME equipped with a tunable single line filter SuperK VARIA) to characterize the combined SLM-doublet device (Fig. 4 in the main text). For this, we build the BFP imaging setup depicted in Fig. S4A. A collimated laser beam illuminates the sample. Subsequently, the diffraction pattern is collected by an objective. The sample image plane and the objective BFP are imaged by means of additional optics to a CCD camera (Thorlabs CS165MU). An adjustable diaphragm is utilized for the image plane spatial filtering to cut undesirable light transmitted through, out of the active area. Importantly, there is a difference in the measurement arrangement between the SLM only and the SLM-doublet system. Fig. S4B illustrates the positions of the image plane for both cases. In the latter, the image plane is located inside the second lens of the doublet. Intuitively, it can be understood from the ray tracing diagram. Indeed, drawing the extension of the output rays inside the doublet, one can find the origin plane. The bottom panel of Fig. S4B depicts the image plane for the SLM only (the left one) and for the SLM-doublet (the right one) for the binary grating phase profile with 4 electrodes per phase level. One can clearly see that the electrode image is squeezed along one direction due to the cylindrical nature of the doublet. Since the image is demagnified, the spatial filter size should be decreased accordingly. This, in turn, leads to the enhanced beam divergence observed in Fig. 4B of the main text. Since the microlens is cylindrical, one can expect the increased beam divergence to occur along one direction only, so that the far field spot is elongated. However, this is not the case in our BFP measurements due to the circular shape of the spatial filter. Note, however, that the aforementioned elongated spots are observed in Supplementary Movie 3, where no spatial filtering is present.

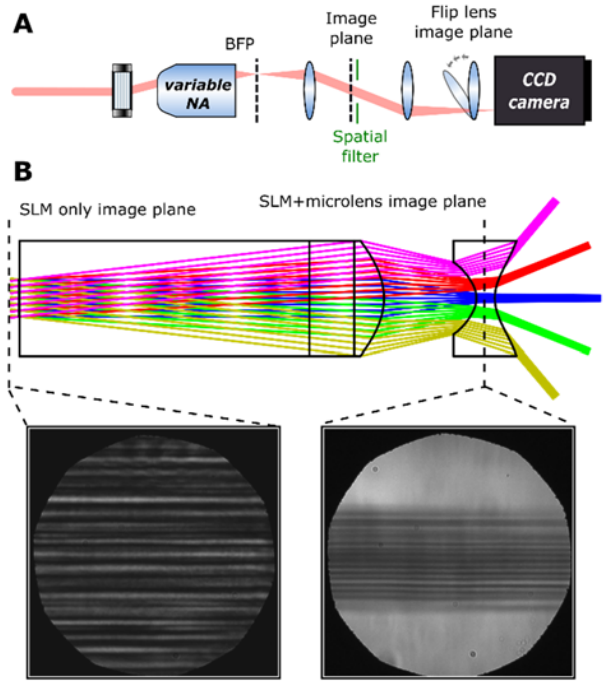

**Fig. S4.**

**Optical characterization of the SLM combined with the doublet microlens.** (A) Schematic of the BFP imaging experimental setup. (B) The top panel depicts the ray tracing diagram together with image plane positions for the SLM case only and SLM-microlens configuration. The bottom panel shows the corresponding images for the binary grating phase profile with 4 electrodes per phase level.

**Movie S1.**

**Switching of the metasurface SLM electrodes.** The movie shows the addressing of each individual metasurface electrode, captured by an optical microscope. To increase the contrast, cross polarizer-analyzer detection is implemented. For each electrode, 3V is applied.

**Movie S2.**

**The real-time beam steering provided by the metasurface SLM**. The movie presents the laser beam steering by alternating the deflection angle for the binary grating regime. The SLM electrodes are programmed to switch the applied grating pitch in ascending order (descending for the diffraction angles). Before each iteration, the electrodes are reset to ensure the correct voltage levels. It manifests as a pure $0^{th}$ order emergence in between of the iterations. In this experiment, the laser is focused to match the size of the active area by a lens with 200 mm focal length. The far-field distribution is visualized on a white paper card.

**Movie S3.**

**The real-time beam steering provided by the combined SLM-doublet device**. The movie presents the laser beam steering for the combined metasurface SLM-doublet configuration. The experimental arrangement is similar to Movie S2. One can clearly see the field of view expansion.